# High temperature heat treatment of B precursor and P.I.T. process optimization to increase $J_c$ performances of MgB$_2$-based conductor

Maurizio Vignolo[1,*], Gianmarco Bovone[2], Cristina Bernini[1], Andrea Palenzona[1,3], Shrikant Kawale[1,2], Gennaro Romano[1] and Antonio Sergio Siri[1,2]

[1] CNR-SPIN, C.so Perrone 24, 16152 Genova, Italy.
[2] University of Genoa, DiFi, Via Dodecaneso 33, 16132 Genova, Italy.
[3] University of Genoa, DCCI, Via Dodecaneso 31, 16132 Genova, Italy.

* corresponding author: Tel +390106598738, e-mail maurizio.vignolo@spin.cnr.it

**Abstract:** Promising results reported in our previous works led us to think that boron's production process plays a crucial role in MgB$_2$ synthesis. A new method for boron preparation has been developed in our laboratory. This particular process is based on magnesiothermic reaction (Moissan's process) with the addition of an initial step that gives boron powder with nano-metric grain size. In this paper we report our efforts regarding optimization of PIT method for these nanometric powders and the resolution of problems previously highlighted such as the difficulty in powder packaging and the high friction phenomena occurring during cold working. This increases cracking during the tape and wire manufacturing leading to its failure. Packaging problems are related to the amorphous nature of boron synthesized in our laboratory, so a crystallization treatment was applied to improve boron's crystallinity. To prevent excessive friction phenomena we synthesized non-stoichiometric MgB$_2$ and using magnesium as lubricant. Our goal is the $J_c$ improvement, but a global physical-chemical characterization was also made to analyze the improvement given by our treatments: this characterization includes X-ray diffraction, $\rho(T)$ measurement, SEM image, besides magnetic and transport $J_c$ measurements.

## 1. Introduction

After 12 years since the discovery of superconductivity in MgB$_2$, many efforts have been done in order to improve its performances and make it competitive with traditional superconductors. This is possible due to its relatively high critical temperature ($T_c$) that permits to use MgB$_2$-based conductors at 20 K by using a cryocooler. Also compared to other superconducting materials MgB$_2$ tapes are easy to manufacture and of low cost. Despite the good results have been achieve so far, huge efforts are necessary especially on $J_c$ improvement, which is the main parameter from application point of view. This could be achieved by obtaining a complex balance between connectivity, upper critical field ($H_{c2}$) and pinning force induced by grain boundaries, precipitates and defects [1]. The realization of MgB$_2$-based conductors following the PIT technique [2] is divided in two alternative techniques: *in-situ* and *ex-situ*. The former one permits a high speed process and high filling factor with low costs [1, 3, 4], but is still difficult to obtain homogeneous long pieces of wire or tape without empty space. The voids are due to the volume reduction that occurs during MgB$_2$ synthesis from magnesium and boron. However our group has faced up mostly the second approach i.e. *ex-situ* [5-7] for the more possibilities in powder manipulation, like milling and milling with dopant, and the possibility of manufacturing homogeneous long length tapes and wires [8]. Recently our group developed a new scalable method to synthesize boron [9] that permits to obtain an improved precursor for MgB$_2$ available at good prices. Main advantage of our process is production of nano-metric particle size boron powder (average grain size 100 nm [9]), which is very important to produce multi-filamentary and thinner conductors [8, 10, 11]. Recently by Brunauer-Emmet-Teller (BET) analysis an average grain size of 60 nm has been confirmed. Despite good and competitive performance obtained, we came across some problems related to the amorphous nature of synthesized boron powder which causes the difficulty in powder packaging. Also the friction phenomena occurred during cold rolling limited grain connectivity and caused failure and cracking of nickel tube, 2 mm in wall thickness with inner diameter of 8 mm, generally used by our group [12, 13]. In order to prevent these problems, we introduced a thermal heat treatment of crystallization at different temperatures and, according to previous works, the optimal temperature seem to be set between 1200 °C and 1900 °C [14-16]. These two temperatures are related to two different crystallization phenomena as the nucleation and grain growth. The first is facilitated at 1200 °C and the second at 1900 °C [17]. Unfortunately we were not able to reach 1900 °C, so we decided to effect the heat treatment at 1700 °C. At that temperature the grain growth is still predominant crystallization phenomenon, but its growth rate is not the highest. Our goal is to reach, at least, crystalline degree of commercial amorphous boron (H. C. Starck grade I), which is not totally amorphous, but presents a polycrystalline nature [15]. This should improve grains connectivity during the manufacturing of conductor, with consequent improvement of $J_c$ and lowering the sheath failure. This unwanted effect was reduced by improving nickel tube toughness by introducing a 300 µm iron sheet between metal tube and MgB$_2$ powder. This Fe layer also serve as a chemical barrier between nickel sheet and MgB$_2$, avoiding formation any intermediate reaction layer. Furthermore, MgB$_2$ was synthesized with 5 wt% magnesium deficiency. Lacking Mg (to complete the stoichiometry) was added during the tube filling process to act as lubricant. By applying these treatments and precautions, we noticed a considerable reduction of nucleation and propagation of cracks during last steps of conductor's manufacturing. In addition a comparison between boron synthesized in our laboratory and commercial one was given to evaluate the real effects of these treatments, besides a comparison of previous samples of milled MgB$_2$ from commercial boron (CB) [18, 19] and previous results of MgB$_2$ from laboratory made boron (LB) [9].

## 2. Experimental

*2.1. Boron powder preparation and characterization.*

Details of boron synthesis and purification have been given in [9]. Once boron is purified at the end of the process two amounts of powder were taken from the same batch to effect the heat treatment of crystallization. The first one (12LB) has been set in an iron boat and placed in a tubular furnace at 1200 °C for 120 minutes, under Ar flow. The second sample (17LB) was put in tantalum crucible, sealed in a quartz ampoule, and heat treated at 1700 °C through an inductive furnace, for 30 minutes. X-ray analysis was made on these powders at the end of the treatment, on pristine powder (LB), and on commercial boron (CB), to verify the improvement of crystal quality led by heat treatment. XRD measurements were performed using a PHILIPS (PW 1170/90 Cu Kα) diffractometer at 40 KV and 30 mA.

*2.2. $MgB_2$ bulk, powder and wire preparation and their characterization.*

After thermal treatment the powders are used to synthesize stoichiometric $MgB_2$ bulks [20] (AP12LB and AP17LB) to verify if this has led any improvement to magnetic $J_c$ and grain connectivity without compromising $T_c$. $T_c$ values were evaluated by measuring the drop of resistivity cooling down the sample and grain connectivity was studied by calculating the difference between $\rho(300\ K)$ and $\rho(40\ K)$ [21]. Quantum Design Physical Properties Measurement System (PPMS) up to 9 T was employed to evaluate $H_{c2}$ in bulk samples, using a current of 1 mA with the magnetic field applied perpendicular to the samples surface. $H_{c2}$ values were taken at 90% of the resistive transition and $H^*$ at 10% [22]. The remaining part of boron powder was used to synthesize $MgB_2$ conductors by *ex-situ* PIT technique [12, 13, 18]. The details of this are given in [23] in short synthesis temperature of 920 °C under Ar-$H_2$ flow in an iron crucible with a niobium inner sheet. In present work the synthesis was carried out with 5 wt% deficiency of Mg, to get a non-stoichiometric superconducting powder. The deficiency of Mg is restored during tube filling process, which also serve as lubricant, preventing the troublesome friction phenomena occurring between grains of $MgB_2$ synthesized with laboratory made boron. The final dimensions of so prepared conductor was a square section of 1.4 mm x 1.4 mm. Wires were cut in different lengths to be able to measure in different instruments, and sintered at 920 °C for 18 min in a tubular furnace under Ar-$H_2$ flow. Only one sample has been sintered at 660 °C for 2880 min to verify if different sintering processes influence $J_c$ performances. Both magnetic and transport $J_c$ measurements were made. Magnetic $J_c$ was evaluated in a 6 mm length wire, by applying Bean's Model [24] to *M-H* hysteresis loop by a commercial 5.5 T MPMS Quantum Design Squid Magnetometer. Transport critical current ($I_c$) measurements at 4.2 K were performed both at Grenoble High Magnetic Field Laboratory (GHMFL) and at the Physics Department of Genoa University (DiFi), respectively on 9 cm long sample up to 13 T and on 20 cm long sample up to 7 T. The criteria adopted to determine $I_c$ was 1 μV/cm and $I_c$ data were converted into $J_c$ values by SEM image. The samples of boron and

MgB$_2$ used for various measurements, reported in this paper, are summarized in Table 1 and 2 respectively.

## 3. Results and discussion

*3.1. Characterization of boron samples.*

Fig.1 shows the XRD patterns of various boron powders previously discussed: amorphous (LB) and heat treated powder (12LB, 17LB) are compared with commercial boron (CB, as purchased). Remarkable difference in crystal quality between LB and CB was observed. Boron synthesized by our group is amorphous, while the purchased one shows a polycrystalline nature. A so huge difference in crystal quality is strictly correlated with the different grain size of the two powders, like already reported in Ref [9]. It is also necessary to take into account that if the grain size is comparable to the crystallite size a large presence of grain boundaries occurs, as consequence a larger amount of defect have a detrimental effect on the crystal quality of the corresponding sample. As told above, this is a troublesome problem because our experience in conductor's manufacturing is based on CB powders, while we found difficulties in adapting the process on this new LB powder. Hence to improve crystal level of LB powder we applied heat treatments at 1200 and 1700 °C. As shown in Figure 1, although these heat treatments did not allow us to reach the crystal quality of CB, but it is possible to see a beginning of improvement in the sample heat treated at 1200 °C for 2 hours (12LB). This can be seen by the partial separation of peaks around 20 degree, and by the lowering of amorphous signal at low angles. The sample heat treated at 1700 °C shows an evident contamination during the heating in inductive furnace. At the end of thermal process, boron powder was found to be connected to the tantalum crucible and this can explains the presence of unidentified secondary phases in the XRD pattern. The most intense peaks between 30 and 40 degree seem to be due to a mixture of several tantalum borides, such as B$_2$Ta, BTa and BTa$_2$ [25, 26]. It is not possible to identify the real phase in this 17LB powder, because the most intense peaks of these three boride compounds are about the same angles. It's worth mentioning here that these heat treatments not only preserve the main characteristic of nano-metric particle size [14] of LB, but also found to be useful for avoiding cracking problems during conductor's manufacturing.

*3.2. Characterization of MgB$_2$ bulk samples.*

Bulk samples were prepared with these powders and used to verify effects led by heat treatments of crystallisation on $J_c$, $T_c$ and grain connectivity. Fig. 2 shows magnetic $J_c$ at 5 K (filled symbols) and 20 K (open symbols) of MgB$_2$ bulks synthesized from 12LB (square) and 17LB (circle). As expected from the X-ray patterns on boron powders, the $J_c$ trend confirms that the boron heat treated at 1700 °C was contaminated. At 5K its $J_c$ curve lies parallel under the $J_c$ curve of MgB$_2$ derived from 12LB. Especially for the latter sample, it should be noted that the behaviour is not so different from the performances of the pure samples reported in literature by several groups [27-31]. Further analysis on bulk from 17LB shown a low $T_c$, a low grain connectivity and high impurity degree, and considering the difficulty of the heat treatment itself, with respect the heat treatment at 1200 °C, we decided to continue only with the characterisations of MgB$_2$ from 12LB. Figure 3 reports the measurement of resistivity as function of temperature for MgB$_2$ bulks, one from 12LB (polycrystalline) and the other from LB (amorphous boron not heat treated). The sample AP12LB presents more connected grains by showing a lower $\Delta\rho$ (defined as the difference between $\rho(300\ K)$ and $\rho(40\ K)$) with respect to APLB. As shown in inset of Fig. 3, AP12LB sample has a lower $T_c$ value, probably due to oxidation during the heat treatment from oxygen-impurities in technical argon gas, or from the iron boat. Further improvement shall be done by using highly pure Ar and tantalum boats for future experiments. Fig. 4 shows the $\rho(T)$ behaviour taking into account the effective cross-section area fraction ($A_F$), calculated by Rowell's analysis [32], which results are reported in Table 3. Corrected resistivity has been recalculated using the $A_F$ value instead of the apparent cross section area measured by a micrometer sleeve like used for resistivity data of Fig. 3. Along with samples discussed above, we have also reported data from old samples, prepared in similar conditions using different commercial boron (amorphous boron from Alfa-Aesar for MgB-TS [34]; crystalline boron from Alfa-Aesar for MgB-1S [34]; commercial $^{11}$B with high purity degree (99.5%) for MgB11-1S) [34]. Samples NM and M [22] are bulks produced removing the Ni sheath from the tapes. The boron precursor is the same of CB (H.C. Starck, Grade I), and MgB$_2$ powder were used to fill directly the nickel tube (NM), while for M sample was previously milled. The $A_F$ has been calculated using the formula: $A_F = \Delta\rho_{ideal}/\Delta\rho_{experimental}$, where $\Delta\rho_{ideal}$ corresponds to 4.3 $\mu\Omega$ cm of the single crystal, as suggested by Rowell. If we use 7.3 $\mu\Omega$ cm as the ideal $\Delta\rho$ [33], the corresponding $A_F$ value will be doubled. Rowell's analysis suggests that, the heat treatment of crystallization has improved the effective current-carrying area, by simply transforming amorphous boron precursor into polycrystalline one. NM sample has a very different $A_F$ value, due to a different preparation process. However the milling process introduces amorphization and local defects [35], which causes considerable reduction in $A_F$, noting the correlation between crystalline degree and grain connectivity. The same trend was found for other samples (from MgB-TS to MgB11-1S) [34] where, increasing the crystalline degree and purity $A_F$ values also increases. In order to evaluate irreversibility field ($H^*(0)$) and upper critical field ($H_{c2}(0)$), the dependence of resistivity as a function of temperature at various magnetic fields has been studied in APLB sample and reported in Fig. 5. Characterization of the crystallized AP12LB seems not representative, due to its contamination, but in future, applying more controlled process, it will be possible. In Fig. 6, $H^*(0)$ and $H_{c2}(0)$ have been extrapolated, and show values of 11.2 T for $H^*(0)$ and 21 T for $H_{c2}(0)$. The $H_{c2}(0)$ value is comparable to the corresponding value for M sample

[22] and higher with respect to NM sample. On the other hand $H^*(0)$ value is similar to NM sample, due to a wider in-field superconducting transition of APLB, correlated to a worse grain connectivity and purity degree of boron precursor (respect to NM).

*3.3. Characterization of $MgB_2$ powders and wires.*

Fig. 7 reports XRD patterns of $MgB_2$ samples, synthesized from three different boron precursors (CB, LB and 12LB) to prepare wires. From these patterns you can see that, despite what told above, MgB-LB is more affected by contamination than MgB-CB and MgB-12LB, which have similar patterns. MgB-LB presents residual magnesium and a higher MgO concentration. This contamination is residual from precursor boron, which was not heat treated. A high temperature heat treatment can purify boron from $B_2O_3$, and this is probably the reason for the lower $B_2O_3$ impurity degree in MgB-12LB. However it is not easy to say if the MgO is already present before the measurement or its formation is due to the air exposure during the measurements. Clearly finer the $MgB_2$ powder, higher the surface area for oxidation. Furthermore amorphous boron is more reactive than the polycrystalline one. For all the three samples no secondary phases were recognised, except MgO and a little amount of magnesium in MgB-LB. In Figure 8 the SEM images of different $MgB_2$ powders are presented in order to give some information on the morphology and the grain size. Comparing Fig. 8(a) with 8(b) it is confirmed that the heat treatment on LB at 1200 °C for 2 hours does not change the grain size. The MgB-12LB sample shows higher crystal quality and regular grains than MgB-LB. No changes were found in the MgB-CB samples (fig 8(c) and 8(d)). Fig. 9 shows the transport $J_c$ behaviour of improved $MgB_2$ wires (MgB-12LB and MgB-CB). Improved wires (hollow symbols) are compared with old samples (full symbols) presented in previous paper [9]. It becomes clear that: the decrease in reaction layer thickness, the elemental Mg addition to complete the reaction stoichiometry and the crystallization process on LB powder make it possible to improve $J_c$ by one order of magnitude at 7 T in MgB-12LB, respect to MgB-LB. In order to evaluate accuracy and validity of the instruments and mathematical models used in this work, we reported in Fig. 10 some measurements of the same mono-core wire (with iron barrier and filled with MgB-12LB) carried out with different instruments at different facilities. The full square symbols is referred to the magnetic $J_c$ at 5 K, the cross-shaped square represents the $J_c$ measurement performed at the GHMFL laboratories of Grenoble and the half-full/half-hollow is the $J_c$ measurements performed at DiFi in Genova. The different measurements on diverse pieces of the same sample are in good agreement. A wire sintered at 660 °C for 2880 min (MgB-12LB II) is also reported in Fig. 10 (hollow square symbol) with the intention to investigate the possibility of using laboratory made boron for future developments regarding the *in-situ* and *ex-situ* multi-filamentary wires. The sintering process at 660 °C for 2880 min seems equivalent to the standard process (920 °C for 18 min). This long sintering process has been thought to avoid the diffusion reaction of the magnesium in the Ni sheath and in future could be applied to the multi-filamentary wires, where the iron barrier is not used. In conclusion of our work we compare in Fig. 11 the best wire sample of this work with the best old conductor samples produced in our laboratory. These old samples were prepared using $MgB_2$ milled powders where B precursor was always the CB precursor [18, 19]. Measurements were made at GHMFL at 4.2 K in magnetic field up to 13 T. As it's easily visible this new wire has better performances than the milled at 180 rpm for 72 hrs (6 - 7 T). Between 6 and 8 T its behaviour is really similar to the sample milled for 144 hrs. These results are very promising, because our synthesis procedure is simple, low cost and easily scalable to industrial level with respect to the milling process which need Ar atmosphere and expensive tools made up of WC.

## 4. Conclusions

In our previous work, a new process of boron production useful for the $MgB_2$ synthesis was described. In present paper we report the improvement of that $MgB_2$-based conductor using the same lab-made boron (LB) powder with extra heat treatment which enhanced its crystal quality. Also during wire fabrication an iron sheet has been inserted between Ni sheet and $MgB_2$ powder and some amount of Mg powder was added on latter stages of fabrication procedure, which serve as a lubricant and also complete $MgB_2$ stoichiometry. These changes remarkably enhanced $J_c$ value. With this process the grain size of the $MgB_2$ synthesized by using lab-made boron is about 350 nm (estimated by BET analysis), independently from its amorphous or crystalline nature. The grain size of LB and 12LB was estimated to be about 60 nm (BET analysis). The average grain size of milled powders is 400 nm [18, 22], very close to the grain size of the powders studied in this work. Its well-known that finer precursor powders are always an added advantage on final performance of superconductors, regardless of the process used *in-situ* or *ex-situ*. In fact, more fine the precursor powders more chemically homogeneous the final superconducting product. Furthermore, very fine $MgB_2$ grain size plays a crucial role in the development of the multi-filamentary tapes [8, 10, 11]. The interesting result obtained in this improved $MgB_2$ powder is that it has the same $J_c$ value of tape containing un-doped milled $MgB_2$ powder. This work represents a good starting point for future samples that can be C-doped following the new boron synthesis process [9]. Considering that the comparison in Fig. 11 is between several tapes and a wire we hope to obtain doped samples with $J_c$ behaviour higher than the doped and milled tapes. It's worth to note that the new process is really cheap and if scaled-up on industrial level, abundant amount of B powders can be produced in single batch.

**Acknowledgments**

Part of this work has been supported by EuroMagNET II under the EU contract number 228043. This work is also partially supported by Fondazione Carige. Authors wish to thank Dr. Matteo Tropeano for his PPMS measurements at DiFi department.

**List of tables:**

**Table 1.** Boron powder samples.

| Boron powder samples | source | heat treatment at T (°C) | state |
|---|---|---|---|
| LB | Lab Made | none | amorphous |
| 12LB | Lab Made | 1200 | polycrystalline |
| 17LB | Lab Made | 1700 | polycrystalline |
| CB | Commercial | none | polycrystalline |

**Table 2.** $MgB_2$ samples.

| Sample name | precursor | form | synthesis T (°C) & t (min) | sintering T (°C) & t (min) | source |
|---|---|---|---|---|---|
| APLB | LB | bulk | 950 & 7200 | none | Lab made |
| AP12LB | 12LB | bulk | 950 & 7200 | none | Lab made |
| AP17LB | 17LB | bulk | 950 & 7200 | none | Lab made |
| MgB-LB | LB | powder/wire | 910 & 60 | 920 & 18 | Lab made |
| MgB-12LB | 12LB | powder/wire | 910 & 60 | 920 & 18 | Lab made |
| MgB-17LB | 17LB | powder | 910 & 60 | 920 & 18 | Lab made |
| MgB-12LB II | 12LB | powder/wire | 910 & 60 | 660 & 2880 | Lab made |
| MgB-CB | CB | powder/wire | 910 & 60 | 920 & 18 | Lab made |
| MgB-CB old | CB | powder/wire | 910 & 60 | 920 & 18 | Lab made |

**Table 3.** Data for MgB$_2$ bulks obtained from different boron precursors.

| MgB$_2$ bulk from: | $T_c$ (K) | $\Delta T_c$ (K) | $\rho$(300 K) ($\mu\Omega$ cm) | $\rho$(40 K) ($\mu\Omega$ cm) | $\Delta\rho$ ($\mu\Omega$ cm) | RRR | $A_F$% | $\rho$(0) ($\mu\Omega$ cm) |
|---|---|---|---|---|---|---|---|---|
| APLB | 37.7 | 1.7 | 49.9 | 15.2 | 34.7 | 3.3 | 12.4 | 1.9 |
| AP12LB | 35.7 | 2.4 | 29.3 | 14.4 | 14.9 | 2.0 | 28.9 | 4.2 |
| MgB11-1S [34] | 38.7 | 0.2 | 8.5 | 0.6 | 7.9 | 15.3 | 54.1 | 0.3 |
| MgB-1S [34] | 38.9 | 0.3 | 15.0 | 2.1 | 12.9 | 7.1 | 33.4 | 0.7 |
| MgB-TS [34] | 38.7 | 1 | 133.3 | 40.4 | 92.9 | 3.3 | 4.6 | 1.9 |
| NM [22] | 38.6 | 0.4 | 116 | 55 | 61.0 | 2.1 | 7.0 | 3.9 |
| M [22] | 37 | 1.6 | 382 | 289 | 93.0 | 1.3 | 4.6 | 13.4 |

**Figure captions:**

**Fig. 1.** Comparison of XRD patterns corresponding to different B powders.

**Fig. 2.** Magnetic $J_c$ measurements on MgB$_2$ bulks from boron 12LB and 17LB at 5 and 20 K.

**Fig. 3.** Resistivity vs. temperature of APLB and AP12LB bulks.

**Fig. 4.** Corrected resistivity vs. temperature of APLB and AP12LB bulks.

**Fig. 5.** Resistivity vs. temperature at different magnetic fields in APLB bulk.

**Fig. 6.** $H_{c2}$ and $H^*$ vs. T diagram for APLB bulk (respectively 90% and 10% of normal state resistivity from the resistive transition in magnetic field in perpendicular configuration respect to the sample surface).

**Fig. 7.** XRD patterns of MgB$_2$ powders from CB (commercial), LB (amorphous) and 12LB (policrystalline).

**Fig. 8.** SEM images of MgB$_2$ powders: (a) MgB-LB, (b) MgB-12LB, (c) MgB-CB old and (d) MgB-CB.

**Fig. 9.** Transport $J_c$ measurements on wire samples.

**Fig. 10.** Magnetic and transport $J_c$ measurements on sample MgB-12LB.

**Fig. 11.** Transport $J_c$ measurements. Comparison among the best improved wire and the milled samples.

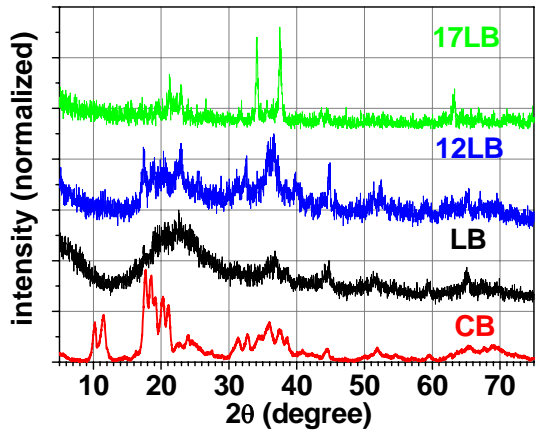

**Fig. 1.** Comparison of XRD patterns corresponding to different B powders.

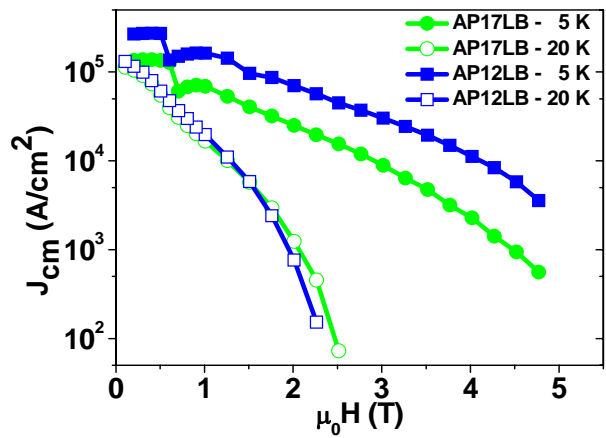

**Fig. 2.** Magnetic $J_c$ measurements on MgB$_2$ bulks using boron 12LB and 17LB at 5 and 20 K.

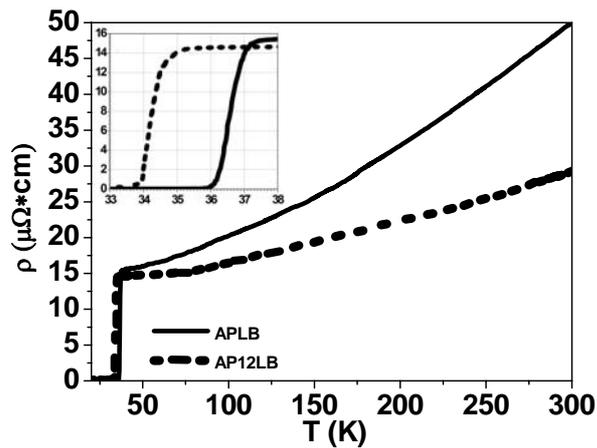

**Fig. 3.** Resistivity vs. temperature of APLB and AP12LB bulks.

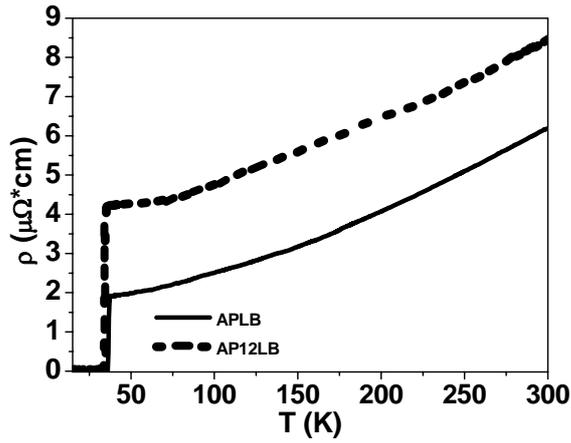

**Fig. 4.** Corrected resistivity vs. temperature of APLB and AP12LB bulks.

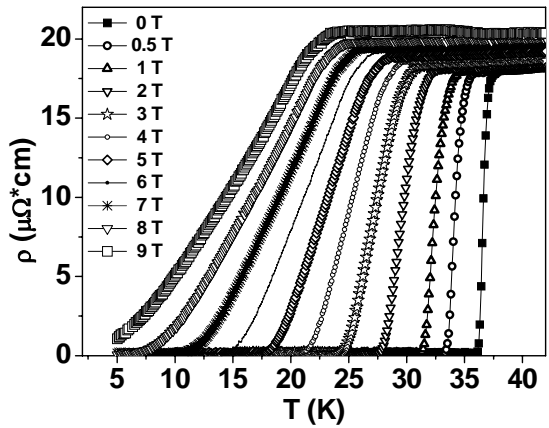

**Fig. 5.** Resistivity vs. temperature at different magnetic fields in APLB bulk.

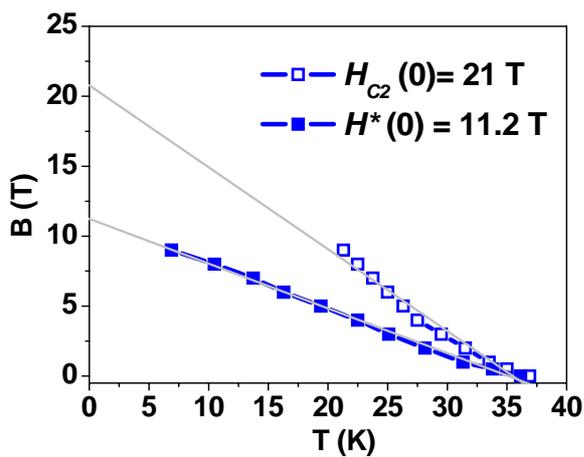

**Fig. 6.** $H_{c2}$ and $H^*$ vs. T diagram of APLB bulk (respectively 90% and 10% of normal state resistivity from the resistive transition in magnetic field in perpendicular configuration respect to the sample surface).

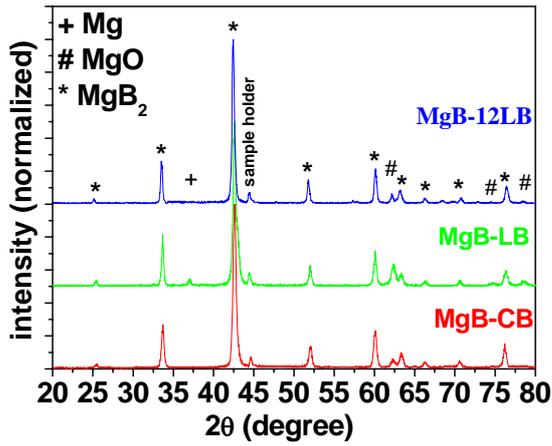

**Fig. 7.** XRD patterns of MgB$_2$ powders from CB (commercial), LB (amorphous) and 12LB (policrystalline).

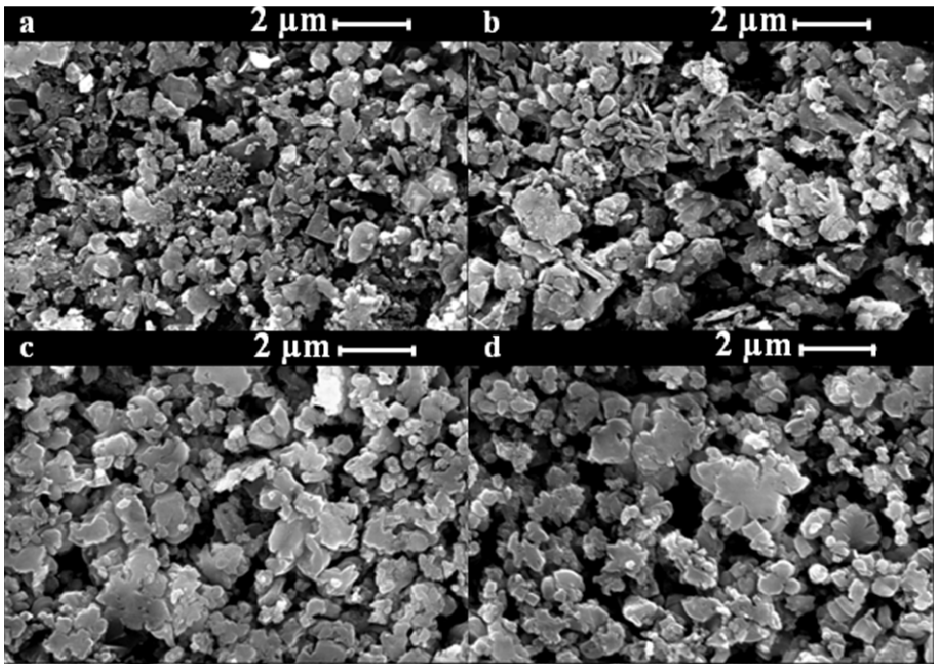

**Fig. 8.** SEM images of MgB$_2$ powders: (a) MgB-LB, (b) MgB-12LB, (c) MgB-CB old and (d) MgB-CB.

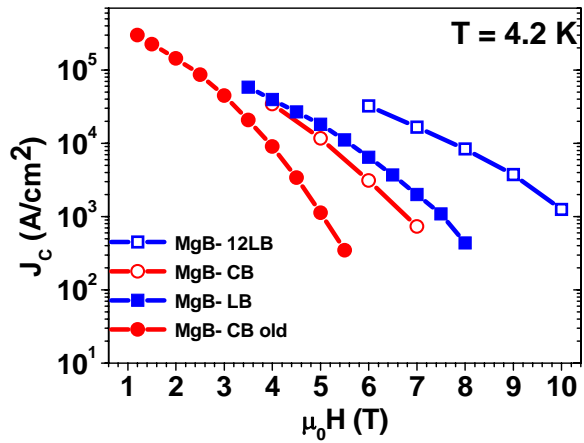

**Fig. 9.** Transport $J_c$ measurements on wire samples.

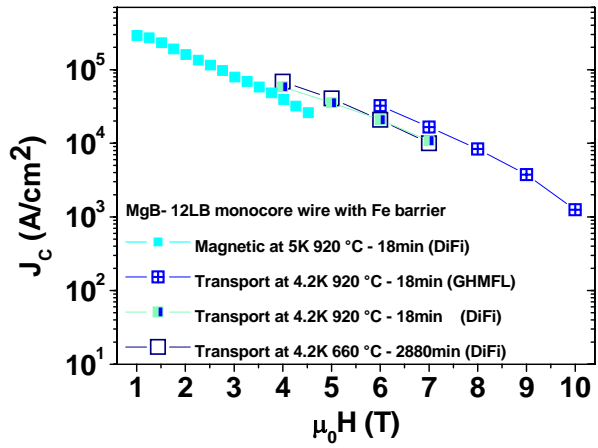

**Fig. 10.** Magnetic and transport $J_c$ measurements on various samples of MgB-12LB.

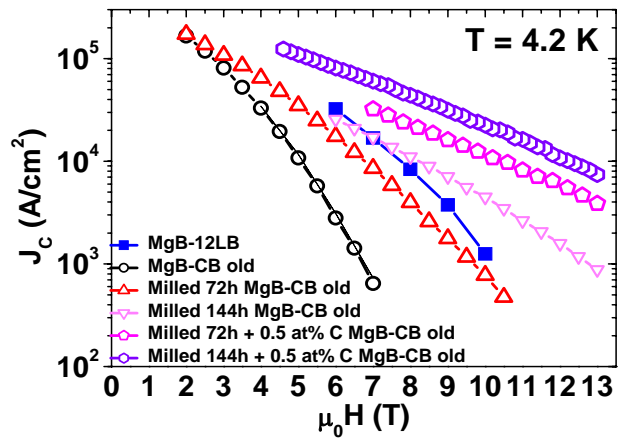

**Fig. 11.** Transport $J_c$ measurements. Comparison among the best improved wire and the samples with milled powders.